\documentclass[prd,aps,preprintnumbers, showpacs, nofootinbib,superscriptaddress, notitlepage, 10pt]{revtex4-1}

\usepackage{epsfig}
\usepackage{bm}
\usepackage{amssymb}
\usepackage{amsmath}
\usepackage{color}
\usepackage{subfigure}
\usepackage[colorlinks,
            linkcolor=blue,
            anchorcolor=black,
            citecolor=blue
            ]{hyperref}

\begin{document}
\title{On the Threshold Resummation in Forward pA Collisions}

\author{Bo-Wen Xiao}
\affiliation{Key Laboratory of Quark and Lepton Physics (MOE) and Institute
of Particle Physics, Central China Normal University, Wuhan 430079, China}

\affiliation{Centre de Physique Th\'eorique, \'Ecole Polytechnique, 
CNRS, Universit\'e Paris-Saclay, Route de Saclay, 91128 Palaiseau, France.}

\author{Feng Yuan}
\affiliation{Nuclear Science Division, Lawrence Berkeley National
Laboratory, Berkeley, CA 94720, USA}

\begin{abstract}
In this paper, using the Higgs production in forward rapidity region in proton-nucleus collisions as an example, we demonstrate that we can construct a systematic formalism for the threshold resummation for forward rapidity particle productions in the saturation formalism. The forward threshold jet function, which satisfies the corresponding renormalization group equation, is introduced into the new factorization formula. This calculation can be easily generalized to other processes, such as single forward hadron productions at forward rapidity region, and have important phenomenological implications. 
\end{abstract}
\maketitle

\section{Introduction}

Single inclusive particle productions in the forward rapidity region in proton-nucleus collisions, $p\, + \, A \to H(y, p_\perp) \, +\, X$, is of particular importance in the search for the onset of the gluon saturation phenomenon, which occurs in high gluon density heavy nucleus target at very small $x$ region. In this process, we measure the transverse momentum distribution of particles produced in the forward rapidity region $y>0$ in the proton beam direction. It is straightforward to find that the active partons with longitudinal momentum fraction $x_p\sim \frac{m_\perp}{\sqrt{s}}e^y$ in the proton projectile lie in the large $x$ region, while the active partons (mostly gluons) with longitudinal momentum fraction $x_A\sim\frac{m_\perp}{\sqrt{s}}e^{-y}$ in the nucleus target are from deeply small-$x$ region when the rapidity $y$ is sufficiently large. Here $m_\perp=\sqrt{m^2+p_\perp^2}$ is defined as the transverse mass of produced particle while $s$ is the center of mass energy. Physically, partons inside the proton projectile, which act as dilute probes, can pick up sizeable amount of transverse momentum, which is of the order of the saturation momentum, after traversing the dense gluonic medium in the heavy nucleus target. Therefore, it is of great interest to study the transverse momentum distribution of particles especially in the low transverse momentum region in order to study the gluon saturation phenomenon. Normally, the so-called dilute-dense factorization, which uses collinear parton distribution functions (PDFs) for partons from the proton side and small-$x$ gluon distributions for the low-$x$ gluon originated from the target nucleus side, is widely adopted to formulate particle productions in the forward region. In this approach, by measuring produced particles in the forward rapidity region, one can take advantage of the extremely asymmetric kinematics (namely, $x_p\to 1$ and $x_A \ll 1$) to maximize the gluon saturation effects. On the other hand, $x_p\to 1$ means that this process is close to the kinematical boundary of the phase space, where the soft gluon radiations become important. The resummation of soft gluon radiations near the kinematical threshold is known as the threshold resummation. The objective of this paper is to understand the threshold resummation in the forward region in the dilute-dense factorization framework. 

There have been great efforts of theoretical and phenomenological studies~\cite{Dumitru:2002qt, Dumitru:2005kb, Albacete:2010bs, Levin:2010dw, Fujii:2011fh, Albacete:2012xq, Albacete:2013ei, Dominguez:2010xd, Altinoluk:2011qy, Chirilli:2011km, Stasto:2013cha, Lappi:2013zma, vanHameren:2014lna, Stasto:2014sea, Altinoluk:2014eka, Watanabe:2015tja, Stasto:2016wrf, Iancu:2016vyg, Ducloue:2016shw, Ducloue:2017dit} on forward hadron productions in $pA$ collisions using the dilute-dense factorization. In particular, the next-to-leading order (NLO) single hadron production cross section in $pA$ collisions in forward rapidity becomes negative in the large transverse momentum region\cite{Stasto:2013cha}. There has been some speculation\cite{threshold} that the threshold resummation could in principle help to mitigate the negativity problem by systematically include higher loop contributions. Furthermore, authors in Ref.~\cite{Stasto:2016wrf} specifically demonstrated that one important source of the negative cross section comes from a term which is proportional to $\alpha_s\ln (1-x_p)$ with $x_p \to 1$. Analytically, one can trace the origin of the $\alpha_s\ln (1-x_p)$ term and find that it is due to the so-called plus distribution in the NLO correction. Sometime, threshold resummation is viewed as a resummation of the `plus' distributions, for example, $\int_\tau^1 \frac{d\xi}{(1-\xi)_+} \sim \ln (1-\tau)$ in the limit $\tau \to 1$, although constant terms associated with soft gluon emissions are also resummed. This implies that the so-called threshold logarithm can start to appear and become important in the dilute-dense factorization approach. In addition, the ultra-forward rapidity hadron spectrum has been measured recently by LHCf\cite{Adriani:2015iwv}, which is sensitive to the very large $x$ region of the projectile hadron and extremely small $x$ region of the target nucleus\cite{Albacete:2016tjq}. In this case, we believe that the threshold resummation becomes indispensable. However, the theoretical framework which can systematically resum such type of threshold logarithms in the dilute-dense factorization approach is not yet available. In the following, we will investigate the the threshold resummation in forward rapidity productions of particles in $pA$ collisions and build such a theoretical framework. 

In general, threshold resummation, which was originally formulated by Sterman\cite{Sterman:1986aj}, and by Catani and Trentadue\cite{Catani:1989ne} for the Drell-Yan process, has been a very important topic in high energy QCD studies for the last thirty years. The resummation technique developed in these two papers has been generalized and applied to many other QCD processes, and has been proved to be very useful in QCD phenomenology. For example, de Florian and Vogelsang\cite{deFlorian:2005yj} applied the threshold resummation to high $p_\perp$ $\pi^0$ productions in $pp$ collisions with rapidity integrated, and found that the resummed results significantly improve the agreement between theoretical predictions and data. Similar technique was also applied to high $p_\perp$ Higgs productions\cite{deFlorian:2005fzc}. In the context of Higgs boson production, the threshold logarithms have also been included in the low transverse momentum resummation, see, for example, Refs.~\cite{Bozzi:2005wk,Bozzi:2007pn}. Furthermore, with the advent of the soft-collinear effective theory (SCET)\cite{Bauer:2000yr}, simple derivation of the factorization formula for the DIS structure function in the threshold limit $x_{\textrm{Bj}}\to 1$ can be achieved as shown in Ref.~\cite{Chay:2005rz, Manohar:2003vb, Becher:2006nr, Becher:2006mr}. In addition, there have been interesting studies on the joint resummation\cite{Li:1998is, Laenen:2000de, Kulesza:2003wn} of transverse momentum logarithms and threshold logarithms, which resemble some similar physical idea as the joint resummation that we are presenting below. The major difference is that we are working in the dilute-dense factorization approach with fixed forward rapidity. 

Let us take the example of the Higgs boson production in gluon-gluon fusion process with an effective Lagrangian~\cite{Dawson:1990zj,Idilbi:2005ni, Idilbi:2006dg} to demonstrate the formulation and factorization. As the main result of this paper, in the limit which the Higgs mass $M$ is much greater than the measured transverse momentum $k_\perp$, the factorization formula for forward rapidity Higgs production in $pA$ collisions can be written as 
\begin{equation}
\frac{1}{\sigma_0}\frac{d\sigma}{dy d^2 k_\perp} =\int_\tau^1 \frac{dx}{x}  \tau g (x, \mu )   \int \frac{d^2 x_\perp d^2x_\perp^\prime}{(2\pi)^2} e^{-i k_\perp \cdot b_\perp } S_{x_A}^{WW}(x_\perp, x_\perp^\prime)  \,  S^{Sud}(M^2, \mu_b^2)\, \Delta(\mu^2, \mu_b^2, \ln \frac{x}{\tau}) \, C(\alpha_s),  \label{fact}
\end{equation}
where $\sigma_0 =\frac{g_\phi^2}{4g^2 (N_c^2-1)}$, $\mu_b^2\equiv\frac{c_0^2}{b_\perp^2}$ with $c_0=2e^{-\gamma_E}$ and $b_\perp\equiv x_\perp -x_\perp^\prime$. Eq.~(\ref{fact}) is a very interesting and elegant formula, which encodes collinear resummation in $g (x, \mu )$, small-$x$ evolution as well as multiple interactions in $S_{x_A}^{WW}(x_\perp, x_\perp^\prime) $\cite{Dominguez:2011gc}, transverse momentum resummation in the conventional Sudakov factor $S^{Sud}(M^2, \mu_b^2)$\cite{Collins:1984kg, Mueller:2013wwa, Balitsky:2015qba, Xiao:2017yya} and the threshold resummation in the new factor, namely the forward threshold jet function $\Delta(\mu^2, \mu_b^2, \ln \frac{x}{\tau})$.\footnote{We name the function $\Delta(\mu^2, \mu_b^2, \ln \frac{x}{\tau})$ as the forward threshold jet function, because it resums threshold logarithms in the forward rapidity region in the dilute-dense factorization formalism, and its analytical form looks very similar to the jet function in SCET.} $C(\alpha_s)$ represents the hard coefficient expanded in terms of $\alpha_s$ without any large logarithms. Essentially, Eq.~(\ref{fact}) also resums the threshold logarithms of type $\alpha_s \ln (1-\tau)$ with $\tau\equiv \frac{M}{\sqrt{s}}e^y \to 1$ in the forward rapidity ($y$) Higgs production. Interestingly, we find that the new forward theshold jet function $\Delta$ satisfies a renormalization group equation (RGE) similar to the RGE first derived by Becher and Neubert\cite{Becher:2006qw, Becher:2006nr} for jet functions in SCET, which can also be derived from the analytical solution of the Dokshitzer-Gribov-Lipatov-Altarelli-Parisi (DGLAP) evolution equation at the end point\cite{Becher:2006mr}. 

When the anomalous dimension $ \gamma_{\mu, b_\perp} \equiv N_c \int_{\mu^2}^{\mu_b^2} \frac{d \mu^{\prime 2}}{\mu^{\prime 2}} \frac{\alpha_s(\mu^{\prime 2})}{\pi} >0$, one obtains $\Delta(\mu^2, \mu_b^2, \ln \frac{x}{\tau})=\frac{e^{\left(\beta_0-\gamma_E\right) \gamma_{\mu, b_\perp} }}{\Gamma [ \gamma_{\mu, b_\perp} ]} z^{\gamma_{\mu, b_\perp} -1} $ at the leading logarithmic level with $z=\ln\frac{x}{\tau}$. Eq.~(\ref{fact}) is insensitive to the choice of factorization scale (also the same as renormalization by choice) $\mu$. If $\mu^2$ is set to be $\mu_b^2$, it is straightforward to show that $\Delta$ becomes $\delta(\ln\frac{x}{\tau})$, which reduces Eq.~(\ref{fact}) to previous results obtained in Ref.~\cite{Mueller:2013wwa}. By setting $\mu^2=\mu_b^2$ in the collinear gluon distribution $g(x, \mu)$, we implicitly encode the threshold logarithms in the collinear gluon distribution by changing its scale with $b_\perp$. This requires the information of collinear PDF $g(x, \mu)$ over a large range of $\mu$ at given $x$. Since PDFs have large uncertainties in the large $x$ and large $\mu^2$ region, it is better to choose $\mu^2$ to be a constant, and perform the threshold resummation explicitly as in Eq.~(\ref{fact}), which is presumably more stable and accurate in the threshold limit. Furthermore, when $b_\perp$ gets large, $\mu_b$ can become smaller than the lowest scale that collinear PDFs are defined. Without a cutoff prescription in the large $b_\perp$ region, this can also be a problem. In addition, in the NLO single hadron production, normally the factorization scale is chosen to be a constant in practice in Ref.~\cite{Stasto:2014sea, Watanabe:2015tja} in order to avoid some technical difficulties in the numerical calculation. In addition, more close connections to the SCET formalism can be established in forward hadron(jet) productions. This will be presented in a separate work with numerical results.

The threshold resummation with respect to the double differential (rapidity and transverse momentum) cross section considered here, as shown in the LHS of Eq.~(\ref{fact}), has been studied previously\cite{Kidonakis:1998nf, Catani:2013vaa, deFlorian:2013qia} in different physical framework and kinematical region. As pointed out in Ref.~\cite{deFlorian:2013qia}, the threshold correction becomes quite large in forward rapidity region $y \sim 4$ due to small-$x$ contributions. In our approach, Eq.~(\ref{fact}) resums both small-$x$ and threshold logarithms, which is in principle more suitable framework to describe forward rapidity particle productions. It would also be very interesting to see the arise of small-$x$ logarithms and corresponding resummations in the forward rapidity production of jets following the framework developed in Ref.~\cite{deFlorian:2013qia}. 

In the following section, we will derive $\Delta(\mu^2, \mu_b^2, \ln \frac{x}{\tau})$ in Eq.~(\ref{fact}) and comment on the application of the threshold resummation in other processes in Sec.II. Before we conclude in Sec. IV, several comments regarding the forward threshold resummations are made in Sec. III.

\section{Forward Threshold Resummation in Higgs Production in pA collisions}
For the sake of simplicity, let us use the forward Higgs production as an instructive example to demonstrate the threshold resummation in saturation formalism. This process is the simplest one since there is no final state gluon radiation. In addition, one can see that the leading power contribution comes from a few diagrams, while the rest of graphs are suppressed by factors of $\frac{k_\perp}{M} \ll 1$. We follow closely the calculation shown in Ref.~\cite{Mueller:2013wwa}, which shows that the LO contribution reads 
\begin{equation}
\frac{d \sigma_{LO}}{d\tau d^2 k_\perp}=\sigma_0 g(\tau, \mu)\int \frac{d^2 x_\perp d^2x_\perp^\prime}{(2\pi)^2} e^{-i k_\perp \cdot b_\perp} S_{x_A}^{WW}(x_\perp, x_\perp^\prime),
\end{equation} 
where $S^{WW}_{x_A}\equiv - \left\langle \textrm{Tr} \left[\partial^i U_{x_\perp}\right] U^\dagger_{x^\prime_\perp} \left[\partial^i U_{x^\prime_\perp}\right] U^\dagger_{x_\perp}\right\rangle_{x_A}$ resums the multiple gluon exchanges between the active gluon and the target nucleus. At LO, the transverse momentum $k_\perp$ of the produced Higgs solely comes from the small-$x$ Weizs\"{a}cker-Williams (WW) gluon distribution represented by the Fourier transform of $S^{WW}_{x_A}$\cite{Dominguez:2010xd}. At one-loop order, working in the leading power $\frac{k_\perp}{M} \ll 1$ limit, we find the following corrections\footnote{Here we assume that the strong coupling is fixed at the moment to illustrate the double and single logarithms. It is straightforward to extend the results to the running couple case for the final results. It is also clear that the Landau pole problem does not appear in this formalism, since the running coupling is determined by the scales between $\mu$ and $\mu_b$. In numerical calculations, $\mu_b$ is always kept larger than $\Lambda_{QCD}$.} 
\begin{eqnarray}
\frac{d \sigma_{1-loop}}{d\tau d^2 k_\perp}&=&  \sigma_0 \int \frac{d^2 x_\perp d^2x_\perp^\prime}{(2\pi)^2} e^{-i k_\perp \cdot b_\perp} S_{x_A}^{WW}(x_\perp, x_\perp^\prime) \frac{\alpha_s N_c}{\pi} g(\tau, \mu) \left[-\frac{1}{2}\ln^2 \frac{M^2}{\mu_b^2}+ \beta_0 \ln \frac{M^2}{\mu_b^2} +\frac{\pi^2}{2} \right] \notag \\
&&+  \sigma_0  \int \frac{d^2 x_\perp d^2x_\perp^\prime}{(2\pi)^2} e^{-i k_\perp \cdot b_\perp} S_{x_A}^{WW}(x_\perp, x_\perp^\prime) \frac{\alpha_s N_c}{\pi}\ln \frac{\mu_b^2}{\mu^2 }  \int_{\tau}^1 \frac{d\xi}{\xi} \mathcal{P}_{gg} (\xi)g\left(\frac{\tau}{\xi}, \mu\right), \label{1loop} \end{eqnarray}
where the terms in the first line of the above equation are conventional Sudakov type logarithms ($\mathcal{O}(\alpha_s\ln^i \frac{M^2}{\mu_b^2} )_{(i=1,2)}$) associated with the transverse momentum resummation\cite{Collins:1984kg}, while the terms in the second line can give rise to the threshold logarithms in the $\tau \to 1$ limit. It is also interesting to note that the latter contribution depends on the splitting function $\mathcal{P}_{gg} (\xi)$, since it is related to the renormalization of the collinear singularity in the gluon PDF. Despite the fact that the Sudakov resummation and the threshold resummation are both resummations with respect to soft gluon contribution, we will distinguish them in our work, since they have different physical interpretation. Here we have defined $\beta_0=\frac{11}{12}- \frac{N_f}{6N_c}$ and 
\begin{equation}
\mathcal{P}_{gg} (\xi)=\frac{\xi}{(1-\xi)_+}+\frac{1-\xi}{\xi}+\xi(1-\xi)+\beta_0\delta(1-\xi).
\end{equation}
Due to the presence of the plus-function and $\delta$ function in the gluon splitting function $\mathcal{P}_{gg} (\xi)$, the end point contributions in the above one-loop corrections, in particular the threshold logarithms, become important when $\tau \to 1$.

The factorization in the threshold resummation can be illustrated and achieved in the Mellin space. Let us define the Mellin transform and inverse Mellin transform as follows
\begin{equation}
f_N =\int_0^1 dx x^{N-1} f(x) \, , \quad  f(x)=\frac{1}{2\pi i}\int_{\mathcal{C}} dN x^{-N} f_N,  \label{mellin}
\end{equation}
where $\mathcal{C}$ represents the properly chosen contour which puts all the integrable poles to its left side. For sufficiently large $N$, the Mellin transform integral is dominated by the end point where $x\sim 1-\frac{1}{N}$. In the Mellin space, the LO cross-section reads
\begin{equation}
\frac{d \sigma^N_{LO}}{ d^2 k_\perp} =\int_0^1 d\tau \tau^{N-1} \frac{d \sigma_{LO}}{d\tau d^2 k_\perp}=\sigma_0 g_N \int \frac{d^2 x_\perp d^2x_\perp^\prime}{(2\pi)^2} e^{-i k_\perp \cdot b_\perp} S_{x_A}^{WW}(x_\perp, x_\perp^\prime),
\end{equation}
where $g_N\equiv \int_0^1 d\tau \tau^{N-1} g(\tau) $. The first line of the 1-loop result in Eq.~(\ref{1loop}) can be transformed similarly. The second line can be transformed as follows
\begin{eqnarray}
\int_0^1d\tau  \tau^{N-1}\int_{\tau}^1 \frac{d\xi}{\xi} \mathcal{P}_{gg} (\xi)g\left(\frac{\tau}{\xi}, \mu\right) 
 = \int_0^1d\tau  \tau^{N-1}\int_{0}^1d\xi \int_0^1 dx \delta (\tau - x \xi )\mathcal{P}_{gg} (\xi)g\left(x, \mu\right) 
   =\mathcal{P}_{gg}(N) g_N ,
   \end{eqnarray}
where one can find
\begin{equation}
\mathcal{P}_{gg}(N) =\int_0^1 d\xi \xi^{N-1} \mathcal{P}_{gg} (\xi) = -\gamma_E -\psi(N)+\beta_0 +\frac{2}{N(N^2-1)}-\frac{1}{N+2},   
\end{equation} 
where $\psi(N)$ is the digamma function. In the large $N$ limit, $\psi(N)$ has the following asymptotic expansion
\begin{equation}
\psi(N) = \ln N-\frac{1}{2N} -\sum_{n=1}^\infty \frac{B_{2n}}{2n N^{2n}}, \label{psiexpansion}
\end{equation}
where $B_{2n}$ is the $2n$-th Bernoulli number. To proceed, we write $\mathcal{P}_{gg}(N) \simeq  -\ln N-\gamma_E+\beta_0$ where higher order terms $\mathcal{O}\left(\frac{1}{N}\right)$ have been neglected. 
Because the resummation of threshold logarithms and Sudakov logarithms are with respect to soft gluons radiations which always factorize, it is expected that $\mathcal{P}_{gg}(N)$ exponentiates when arbitrary number of soft gluon emissions is resummed\cite{Gatheral:1983cz, Frenkel:1984pz}. In the Appendix~\ref{a1}, an alternative derivation of our results following the RGE idea in Ref.~\cite{Becher:2006mr} is also provided.
Therefore, we can arrive at the following resummed formula in the Mellin space
\begin{eqnarray} 
\frac{d \sigma^N_{Res}}{ d^2 k_\perp}&=& \sigma_0  \int \frac{d^2 x_\perp d^2x_\perp^\prime}{(2\pi)^2} e^{-i k_\perp \cdot b_\perp} S_{x_A}^{WW}(x_\perp, x_\perp^\prime)  \exp \left[\frac{\alpha_s N_c}{\pi}\left(-\frac{1}{2}\ln^2 \frac{M^2}{\mu_b^2}+ \beta_0 \ln \frac{M^2}{\mu_b^2} \right) \right]   \notag \\
&&\times g_N (\mu )\exp\left[- \frac{\alpha_s N_c}{\pi}\ln \frac{\mu_b^2}{\mu^2 }\left(\ln N+\gamma_E-\beta_0\right)\right]  \left[1+\frac{\alpha_s}{\pi}\frac{\pi^2}{2}N_c\right].  
\end{eqnarray}
Similar results have been obtained previously in the context of collinear calculations where both transverse momentum and threshold resummations are considered~\cite{Bozzi:2005wk,Bozzi:2007pn}. In the following, we provide a derivation in terms of the inverse Mellin transform for the logarithms associated with the threshold resummation in the above equation. For example, 
by performing the inverse Mellin transform, the final resummation formula can be cast into 
\begin{eqnarray} 
\frac{d \sigma_{Res}}{ d\tau d^2 k_\perp}&=& \sigma_0  \int \frac{d^2 x_\perp d^2x_\perp^\prime}{(2\pi)^2} e^{-i k_\perp \cdot b_\perp} S_{x_A}^{WW}(x_\perp, x_\perp^\prime)  \exp \left[\frac{\alpha_s N_c}{\pi}\left(-\frac{1}{2}\ln^2 \frac{M^2}{\mu_b^2}+ \beta_0 \ln \frac{M^2}{\mu_b^2} \right) \right] \notag \\
&&\times\int_\mathcal{C} \frac{dN}{2\pi i}  \tau^{-N} g_N (\mu )\exp\left[- \gamma_{\mu, b_\perp} \ln \frac{Ne^{\gamma_E}}{e^{\beta_0}}\right] \left[1+\frac{\alpha_s}{\pi}\frac{\pi^2}{2}N_c\right]. \label{res2}  
\end{eqnarray}
It is interesting to note that the inverse Mellin transform in Eq.~(\ref{res2}) can be performed analytically by applying the following
identity~\cite{Catani:1996yz} ,
\begin{eqnarray}
 \int_\mathcal{C} \frac{dN}{2\pi i}  \left(\frac{x}{\tau}\right)^{N} \exp\left[- \gamma \ln N \right]=\int_\mathcal{C} \frac{dN}{2\pi i}  \left(\frac{x}{\tau}\right)^{N} N^{-\gamma}= \frac{\theta(x-\tau)}{\Gamma(\gamma)} \left[\ln x -\ln \tau \right]^{\gamma -1}, \quad \textrm{Re} [\gamma ] >0 \ ,
\end{eqnarray}
which is derived by integrating above and below its branch cut. Analogous to the analytical continuation of the gamma function, the above identity can be extended to the full complex plane. With this identity, the resummation result can be written as
\begin{eqnarray} 
\frac{d \sigma_{Res}}{ d\tau d^2 k_\perp}&=& \sigma_0  \int \frac{d^2 x_\perp d^2x_\perp^\prime}{(2\pi)^2} e^{-i k_\perp \cdot b_\perp} S_{x_A}^{WW}(x_\perp, x_\perp^\prime) \notag \\
&& \times  \exp \left[\frac{\alpha_s N_c}{\pi}\left(-\frac{1}{2}\ln^2 \frac{M^2}{\mu_b^2}+ \beta_0 \ln \frac{M^2}{\mu_b^2} \right) \right]  \left[1+\frac{\alpha_s}{\pi}\frac{\pi^2}{2}N_c\right] \notag \\
&&\times \frac{e^{\left(\beta_0-\gamma_E\right) \gamma_{\mu, b_\perp} }}{\Gamma [ \gamma_{\mu, b_\perp} ]}\int_\tau^1 \frac{dx}{x}  g (x, \mu )   \left[\ln x -\ln \tau \right]^{\gamma_{\mu, b_\perp} -1}, \quad \gamma_{\mu, b_\perp} >0.  \label{res3}
\end{eqnarray}
The resummation result as in Eq.~(\ref{res3}) becomes singular when $\gamma_{\mu, b_\perp} \leq 0$ at $x=\tau$. Nevertheless, we can use the following trick in terms of the analytical continuation to extend to the region $-1< \gamma_{\mu, b_\perp}\leq 0$
\begin{eqnarray}
&&\int_0^1 \frac{dx}{x} g(x, \mu) \int_\mathcal{C} \frac{dN}{2\pi i}  \left(\frac{x}{\tau}\right)^{N} \exp\left[- \gamma_{\mu, b_\perp}\ln N \right]\notag \\
&=& \int_0^1 \frac{dx}{x}  \left[g(x, \mu)-g(\tau, \mu)+g(\tau, \mu)\right] \int_\mathcal{C} \frac{dN}{2\pi i}  \left(\frac{x}{\tau}\right)^{N} \exp\left[- \gamma_{\mu, b_\perp}\ln N \right]  \notag \\
&=& \int_\tau^1 \frac{dx}{x} \left[g(x, \mu)-g(\tau, \mu)\right] \frac{ \left[\ln x -\ln \tau \right]^{\gamma_{\mu, b_\perp} -1}}{\Gamma[\gamma_{\mu, b_\perp}]} +g(\tau, \mu) \frac{ \left[\ln \left(\frac{1}{\tau} \right)\right]^{\gamma_{\mu, b_\perp} }}{\Gamma[\gamma_{\mu, b_\perp}+1]} . \label{gmone}
\end{eqnarray}
In addition, occasionally, it is also useful to evolve PDFs backwards which means $\mu^2 \gg \mu_b^2$.  Then we need to further analytically continue to the region where $\gamma_{\mu, b_\perp} >-2$ as follows
\begin{eqnarray}
&&\int_0^1 \frac{dx}{x} g(x, \mu) \int_\mathcal{C} \frac{dN}{2\pi i}  \left(\frac{x}{\tau}\right)^{N} \exp\left[- \gamma_{\mu, b_\perp}\ln N \right]\notag \\
&=& \int_\tau^1 \frac{dx}{x} \left[g(x) - g(\tau) -\tau g^\prime (\tau)\ln \frac{x}{\tau}\right]\frac{ \left[\ln x -\ln \tau \right]^{\gamma_{\mu, b_\perp} -1}}{\Gamma[\gamma_{\mu, b_\perp}]} 
+ \frac{ \left[\ln \left(\frac{1}{\tau} \right)\right]^{\gamma_{\mu, b_\perp} }}{\Gamma[\gamma_{\mu, b_\perp}]}\left[\frac{g(\tau)}{\gamma_{\mu, b_\perp}} +\frac{\ln \left(\frac{1}{\tau} \right) }{\gamma_{\mu, b_\perp}+1} \tau g^{\prime}(\tau)  \right].\label{gm2}
\end{eqnarray}

In fact, by utilizing the same subtraction method, we can extend of the region of validity for $\gamma_{\mu, b_\perp}$ to any negative value, and even to $-\infty$. At last, in the case of running coupling, the final resummed result can be written as
\begin{eqnarray} 
\frac{d \sigma_{Res}}{ d\tau d^2 k_\perp}&=& \sigma_0  \int \frac{d^2 x_\perp d^2x_\perp^\prime}{(2\pi)^2} e^{-i k_\perp \cdot b_\perp} S_{x_A}^{WW}(x_\perp, x_\perp^\prime) \exp \left[- N_c\int_{\mu^2_b}^{M^2} \frac{d \mu^{\prime 2}}{\mu^{\prime 2}}\frac{\alpha_s (\mu^{\prime 2}) }{\pi} \ln \frac{M^2}{\mu^{\prime 2}} +\beta_0 N_c \int_{\mu_b^2}^{M^2}  \frac{d \mu^{\prime 2}}{\mu^{\prime 2}}\frac{\alpha_s (\mu^{\prime 2}) }{\pi}  \right] \notag \\ 
&&\times e^{(\beta_0-\gamma_E) \gamma_{\mu, b_\perp} }\int_\tau^1 \frac{dx}{x}  g (x, \mu )\frac{ \left[\ln x -\ln \tau \right]_\ast^{\gamma_{\mu, b_\perp} -1} }{\Gamma [ \gamma_{\mu, b_\perp} ]}  \times C(\alpha_s)   , \label{res4}  
\end{eqnarray}
where $ \left[\ln x -\ln \tau \right]_\ast^{\gamma_{\mu, b_\perp} -1}$ is defined in the spirit similar to the so-called star-distribution\cite{Bosch:2004th, Becher:2006mr}. The star-distribution should be understood as the last line of Eq.~(\ref{gmone}) in the region of $-1< \gamma_{\mu, b_\perp}\leq 0$ and as in Eq.~(\ref{gm2}) in the region $\gamma_{\mu, b_\perp}>-2$, which is normally sufficient for the purpose of numerical evaluations.\footnote{Analogous to the plus distribution, the star distribution is defined as follows
\begin{equation}
\int_0^{Q^2} dp^2 \left[\frac{1}{p^2} \left(\frac{p^2}{\mu^2}\right)^\eta \right]_\ast f(p^2) \equiv \int_0^{Q^2} dp^2 \frac{f(p^2) -f(0)}{p^2} \left(\frac{p^2}{\mu^2}\right)^\eta  +\frac{f(0)}{\eta} \left(\frac{Q^2}{\mu^2}\right)^\eta, \notag
\end{equation}
which can be easily shown to be equivalent to the expression used in this paper for the region $-1< \eta \leq 0 $
\begin{eqnarray}
\int_\tau^1 \frac{dx}{x}  g (x)  \frac{\left[\ln x -\ln \tau \right]_\ast^{\eta -1}}{\Gamma [ \eta ]}  
\equiv \int_\tau^1 \frac{dx}{x} \left[g(x)-g(\tau)\right] \frac{ \left[\ln x -\ln \tau \right]^{\eta -1}}{\Gamma[\eta]} +g(\tau) \frac{ \left[\ln \left(\frac{1}{\tau} \right)\right]^{\eta }}{\Gamma[\eta+1]},   \notag 
\end{eqnarray}
if one simply sets $p^2 = \mu^2 \ln \frac{x}{\tau}$ and $Q^2 = \mu^2 \ln \frac{1}{\tau}$. Here $f(p^2)$ and $g(x)$ can be any smooth test functions.} The final resummation result can be written into a compact form by introducing the threshold resummed gluon distribution $ g_{\textrm{t}} (\tau, \mu_b) $ at the scale $\mu_b$
\begin{equation}
 g_{\textrm{t}} (\tau, \mu_b) = \int_\tau^1 \frac{dx}{x}  g (x, \mu ) \Delta(\mu^2, \mu_b^2, \ln \frac{x}{\tau}) 
 = \frac{e^{(\beta_0-\gamma_E) \gamma_{\mu, b_\perp} }}{\Gamma [ \gamma_{\mu, b_\perp} ]}\int_\tau^1 \frac{dx}{x}  g (x, \mu ) \left[\ln \frac{x}{\tau} \right]_\ast^{\gamma_{\mu, b_\perp} -1}, \label{thg}
\end{equation}
where the forward threshold jet function $\Delta(\mu^2, \mu_b^2, \ln \frac{x}{\tau})$ is introduced for the purpose of threshold resummation. 
Plugging above expression into Eq.~(\ref{res4}), we obtain the final result for Higgs boson production in forward $pA$ collisions as described in Eq.~(\ref{fact}) with $C(\alpha_s)= \left[1+\frac{\alpha_s}{\pi}\frac{\pi^2}{2}N_c +\mathcal{O} \left(\alpha_s(1-\tau)\right)\right]$. Since the large $N$ approximation is used in reaching above results, there are corrections of order $\alpha_s(1-\tau)$ which are neglected. The terms which are neglected are explicitly shown in Appendix.~\ref{a2}.

We would like to emphasize that $g_{\textrm{t}} (\tau, \mu_b) $ is independent of the renormalization scale $\mu$ when $\tau$ is sufficiently close to $1$. For quark distributions, similar resummation can be achieved once we replace $N_c$ and $\beta_0$ by $C_F$ and $\frac{3}{4}$, respectively. The off-diagonal channels are suppressed, simply because there are no plus-function or $\delta$-function in the off-diagonal splitting functions. The consequence of the above formula is that it resums important contributions and restores predictive power in the threshold limit. It is not coincidence that the $\mu$ dependence in the collinear PDF $g (x, \mu )$ is offset by that in $\Delta(\mu^2, \mu_b^2, \ln \frac{x}{\tau})$. In fact, the convolution in Eq.~(\ref{thg}) is determined by the end point limit of the DGLAP evolution equation\cite{Becher:2006mr}. Eq.~(\ref{thg}) is a useful formula since it can provide us PDFs at any scale in the threshold limit. Detailed discussion regarding this issue and comparison with previous results are provided in the Appendix~\ref{a1}. As to the threshold logarithms associated with fragmentation functions, an equivalent corresponding equation can be written with respect to fragmentation functions as well. In fact, as far as we know, a simplified version of this formula for valence quarks first appeared in a review paper\cite{Dokshitzer:1978hw} in the beginning of QCD.

\section{Comments on the Forward Threshold Jet Function}

Before we conclude, several comments with respect to the forward threshold jet function $\Delta(\mu^2, \mu_b^2, z=\ln \frac{x}{\tau})$ and Eq.~(\ref{res4}), which is our main result, are in order.
\begin{itemize}
\item With the identity regarding the digamma function $\psi (\gamma)=-\gamma_E +\int_0^1du \frac{1-u^{\gamma -1}}{1-u}$, it is straightforward to check that the forward threshold jet function introduced above 
\begin{equation}
\Delta(\mu^2, \mu_b^2, z=\ln \frac{x}{\tau}) \equiv  \frac{e^{(\beta_0-\gamma_E) \gamma_{\mu, b_\perp} }}{\Gamma [ \gamma_{\mu, b_\perp} ]}   \left[\ln \frac{x}{\tau}\right]^{\gamma_{\mu, b_\perp} -1}, \quad \textrm{when} \quad \gamma_{\mu, b_\perp} >0,
\end{equation} 
is the solution to the following non-local RGE
\begin{equation}
\frac{d\Delta(\mu^2, \mu_b^2, z)}{d\ln \mu}=-\frac{2\alpha_s N_c}{\pi} \left[\ln z\, +\beta_0\right] \Delta(\mu^2, \mu_b^2, z) +\frac{2\alpha_s N_c}{\pi} \int_0^z dz^\prime \frac{\Delta(\mu^2, \mu_b^2, z) - \Delta(\mu^2, \mu_b^2, z^\prime )}{z-z^\prime}. \label{rge}
\end{equation}
It is very interesting to note that the above integro-differential equation almost coincides with the RGE\cite{Becher:2006qw, Becher:2006nr} developed for jet functions in SCET, once we remove the Sudakov type logarithmic terms there and identify $\Gamma_{cusp}=\frac{\alpha_s N_c}{\pi}$ for gluons at one-loop level. To make the connection more manifest, the scale hierarchy $Q\sim M \gg \mu_i \sim \mu \gg \Lambda_{QCD}$ vital to the usual threshold resummation also appears in our calculation once we identify the intermediate scale $\mu_i$ as $\mu_b$, which always shows up in a coordinate space formulation. The form of the RGE in Eq.~(\ref{rge}) is specifically related to the one-loop correction of this particular process. The solution of RGE automatically contains the corresponding resummation. This interesting link with SCET can be useful for us to perform threshold resummations for other processes and go beyond leading logarithmic level with the help of the RGE technique, since $\Gamma_{cusp}$ and $\beta$ functions have been calculated as high as four\cite{Moch:2005ba} and five loops\cite{Herzog:2017ohr}, respectively. 

\item Again when $\gamma_i\equiv N_c \int_{\mu_i^2}^{\mu_{i+1}^2} \frac{d \mu^{\prime 2}}{\mu^{\prime 2}} \frac{\alpha_s(\mu^{\prime 2})}{\pi} >0$, motivated by the discussion in Ref.~\cite{Dokshitzer:1978hw}, we can easily prove that $\Delta(\mu^2, \mu_b^2, \ln \frac{x}{\tau})$ has the following interesting propagation property
\begin{equation}
\int_\tau^1\frac{dx}{x} \Delta(\mu_1^2, \mu_2^2, \ln \frac{1}{x})  \Delta(\mu_2^2, \mu_3^2, \ln \frac{x}{\tau}) =\Delta(\mu_1^2, \mu_3^2, \ln \frac{1}{\tau}).
\end{equation}
This can be interpreted as that the evolution from $1$ to $\tau$ represented by $\Delta(\mu_1^2, \mu_3^2, \ln \frac{1}{\tau})$ can be written as the convolution of two step evolutions after summing over intermediate states. 

\item Let us also explicitly demonstrate that Eq.~(\ref{res4}) resums the threshold logarithms $\alpha_s \ln (1-\tau)$ by assuming $g(x, \mu)=c(\mu) (\ln\frac{1}{x})^{b(\mu)}$ in the $x\to 1$ limit\footnote{This is equivalent to the parametrization of $g(x, \mu)=c(\mu) (1-x)^{b(\mu)}$ at the first order of $(1-x)$ expansion} and $\gamma_{\mu, b_\perp}> 0$ for the sake of simplicity. This allows us to find
\begin{eqnarray} 
\frac{e^{(\beta_0-\gamma_E) \gamma_{\mu, b_\perp} }}{\Gamma [ \gamma_{\mu, b_\perp} ]}\int_\tau^1 \frac{dx}{x}  g (x, \mu ) \left[\ln x -\ln \tau \right]_\ast^{\gamma_{\mu, b_\perp} -1} 
=\frac{e^{(\beta_0-\gamma_E) \gamma_{\mu, b_\perp} } \Gamma [ b(\mu) +1] }{\Gamma [ \gamma_{\mu, b_\perp} +b(\mu) +1]}  g(\tau, \mu) \left(\ln \frac{1}{\tau}\right)^{\gamma_{\mu, b_\perp}}.
\end{eqnarray}
Since $\ln \frac{1}{\tau}=(1-\tau)+\mathcal{O}[(1-\tau)^2]$ in the $\tau \to 1$ limit, we can see that the above expression essentially resums threshold type logarithms schematically. The threshold logarithms can be written as $\gamma_{\mu, b_\perp}  \ln (1-\tau) \sim \alpha_s \ln (1-\tau) $.

\item As a matter of fact, through mathematical induction, one can prove that the formula involving the inverse Mellin transform is valid and well-defined for any value of $\gamma_{\mu, b_\perp}$ in the complex plane
\begin{eqnarray} 
\int_0^1 \frac{dx}{x} g(x, \mu) \int_\mathcal{C} \frac{dN}{2\pi i}  \left(\frac{x}{\tau}\right)^{N} \exp\left[- \gamma_{\mu, b_\perp}\ln N \right] = \sum_{k=0}^\infty \frac{ \left[\ln \left(\frac{1}{\tau} \right)\right]^{\gamma_{\mu, b_\perp} +k }}{\Gamma[\gamma_{\mu, b_\perp}] (\gamma_{\mu, b_\perp}+k)}  g^{(k)}(\tau),\label{sumk}
\end{eqnarray}
with $g^{(k)}(\tau)\equiv \frac{1}{k!} \left.\frac{\partial ^k }{\partial u^k}g(\tau e^u, \mu)\right|_{u=0}$. 
When $ \gamma_{\mu, b_\perp}$ is small, the above series converges quite fast so that the sum of the first two terms is already close to the exact result. However, when $ |\gamma_{\mu, b_\perp}|$ becomes large, the complete summation should be taken into account. 
As expected, for $\gamma_{\mu, b_\perp} >0$, the above summation over $k$ can be easily performed which gives 
\begin{eqnarray} 
\sum_{k=0}^\infty \frac{ \left[\ln \left(\frac{1}{\tau} \right)\right]^{\gamma_{\mu, b_\perp} +k }}{\Gamma[\gamma_{\mu, b_\perp}]}  g^{(k)}(\tau) \int_0^1 du u^{\gamma_{\mu, b_\perp}+k-1}
=\frac{1}{\Gamma [ \gamma_{\mu, b_\perp} ]}\int_\tau^1 \frac{dx}{x}  g (x, \mu ) \left[\ln x -\ln \tau \right]^{\gamma_{\mu, b_\perp} -1}.
\end{eqnarray}
However, when $\gamma_{\mu, b_\perp}$ becomes negative, we can no longer resum all the terms in Eq.~(\ref{sumk}). Instead, the summation over $k$ starts from the first integer value with $k> -\gamma_{\mu, b_\perp} $. This naturally explains the subtraction method employed above and the origin of the star-distributions. For example, from Eq.~(\ref{sumk}), it is now straightforward to find that the following formula gives the star distribution in the region $\gamma_{\mu, b_\perp} > -3$
\begin{eqnarray}
&&\int_0^1 \frac{dx}{x} g(x, \mu) \int_\mathcal{C} \frac{dN}{2\pi i}  \left(\frac{x}{\tau}\right)^{N} \exp\left[- \gamma_{\mu, b_\perp}\ln N \right]\notag \\
&=& \int_\tau^1 \frac{dx}{x} \Delta g^{(3)}(x, \tau, \mu)\frac{ \left[\ln x -\ln \tau \right]^{\gamma_{\mu, b_\perp} -1}}{\Gamma[\gamma_{\mu, b_\perp}]} + \frac{ \left[\ln \left(\frac{1}{\tau} \right)\right]^{\gamma_{\mu, b_\perp} }}{\Gamma[\gamma_{\mu, b_\perp}]}\left[\frac{g(\tau)}{\gamma_{\mu, b_\perp}} +\frac{\ln \left(\frac{1}{\tau} \right) }{\gamma_{\mu, b_\perp}+1} g^{(1)}(\tau) + \frac{\left[\ln \left(\frac{1}{\tau} \right)\right]^2 }{\gamma_{\mu, b_\perp}+2} g^{(2)}(\tau) \right] ,\,\,\,\,\,\,\,\label{gm3}
\end{eqnarray}
where $\Delta g^{(3)}(x, \tau, \mu)=g(x) - g(\tau) -g^{(1)}(\tau) \ln \frac{x}{\tau} -g^{(2)}(\tau) \ln^2 \frac{x}{\tau} $. 

\end{itemize}

\begin{figure}[tbp]
\begin{center}
\includegraphics[width=11cm]{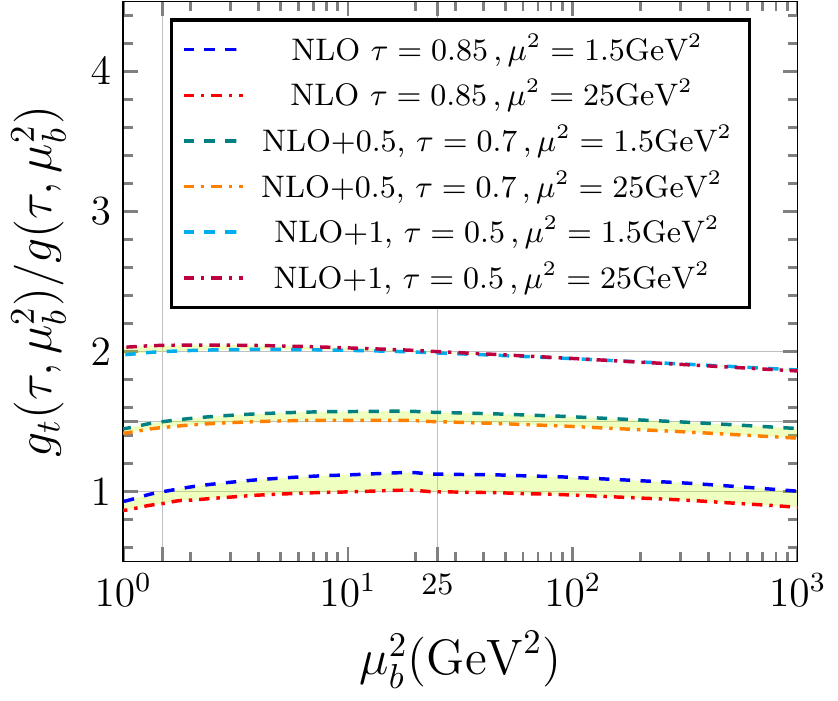} 
\end{center}
\caption[*]{The ratio $\frac{g_{\textrm{t}}(\tau, \mu_b)}{g(\tau, \mu_b)}$ plotted as functions of $\mu_b^2$ for different values of $\tau$ and $\mu^2$ calculated at NLO. All the ratios are rather close to unity. In order to separate the curves for $\tau=0.5$ and $\tau=0.7$ from the $\tau=0.85$ curves, we add $1$ and $0.5$ to the ratio for $\tau=0.5$ and $\tau=0.7$ curves, respectively.}
\label{ratio}
\end{figure}

Finally, we close this section by showing some numeric results for the threshold jet functions at various value of $x$ ($\tau$). In Fig.~\ref{ratio}, we plot the ratio between $g_{\textrm{t}} (\tau, \mu_b)$ computed from Eq.~(\ref{thg}) and $g (\tau, \mu_b)$ at the same scale $\mu_b$ for $\tau =0.5, \, 0.7, \, 0.85$, respectively. We use the MSTW gluon PDF\cite{Martin:2009iq} $g (x, \mu)$ in Eq.~(\ref{thg}) as an input to obtain $g_{\textrm{t}} (\tau, \mu_b)$, and then we use again the MSTW gluon PDF $g (\tau, \mu_b)$ as the denominator. The ratio is expected to be flat and close to unity when $\tau \to 1$, since $g (\tau, \mu_b)$ is the solution to the exact DGLAP equation and $g_{\textrm{t}} (\tau, \mu_b)$ is derived from the endpoint limit of the DGLAP equation. Here, in order to get better numerical agreement, the NLO threshold resummed curves are computed with NLO running coupling and NLO DGLAP equation as given in the end of the Appendix~\ref{a1}. Indeed, we observe that, in the range from $\mu_b^2 =1\textrm{GeV}^2$ to $1000 \textrm{GeV}^2$, $g_{\textrm{t}} (\tau, \mu_b)$ yields agreement with the MSTW gluon PDF at the same scale with roughly less than $10\%$ deviations. Also, we vary the scale $\mu^2$ from $1\textrm{GeV}^2$ to $25\textrm{GeV}^2$, and find that $g_{\textrm{t}} (\tau, \mu_b)$ is insensitive to $\mu^2$ choices as expected. (It seems that the scale dependence gets stronger when $\tau$ gets large. This is due to the fact that the scale evolution gets much more rapid when $\tau$ approaches $1$. Nevertheless, we do see that the curves get more and more flat when $\tau \to 1$.) More interestingly, as shown in the curves with $\mu^2=25\textrm{GeV}^2$, we obtain excellent agreement with the PDF in the region $\mu_b^2 < \mu^2$ as well, where $\gamma_{\mu, b_\perp}$ becomes negative and Eq.~(\ref{gm2}) has to be employed. This proves that the analytical continuation technique works as expected for the so-called backward evolution. Since the MSTW PDFs is provided above $1\textrm{GeV}^2$, we decide to make a cut at $1\textrm{GeV}^2$, although we believe Eq.~(\ref{thg}) can be used to evolve $g (x, \mu)$ from $\mu$ down to the scale $\mu_b$, which can be less than $1\textrm{GeV}^2$. 

We would like to emphasize that the above comparisons help to establish the validity of the threshold jet function, which can be applied to other forward scattering processes. In particular, in forward hadron production in $pA$ collisions, the differential cross section  can be written in terms of the parton distribution at the scale of $\mu_b$, i.e.,  $f^{(a)}(x,\mu_b)$ where $a$ represents a quark or gluon from the incoming nucleon~\cite{Chirilli:2011km}. The threshold resummation can be carried out, by applying the similar technique of this paper. Again, the final result can be cast into $f_t^{(a)}(x,\mu_b)$ multiplied by the resummation result for the hard coefficients, which will be different from the current case. In the current case, the hard part, which only depends on a term proportional to the splitting function and an $\alpha_s$ correction with no threshold logarithms, is quite simple. However, in forward hadron production case, there exist other large threshold logarithms in the hard part, whose resummation will be also important for a precision calculation. We will leave that for a separate publication. 

\section{Conclusion and Outlook}

In this paper, we have demonstrated the resummation of threshold logarithms in the dilute dense factorization which is widely used studying small-$x$ effects in high energy collisions by using complex analysis and RGE methods. The framework discussed above resembles a lot of similarities to the threshold resummation in SCET and traditional resummation in the Mellin space. The advantage of this approach is that final results can be expressed in momentum space analytically with decent numerical accuracy. To obtain more precise results, we believe that we need to adopt the approach derived in Ref.~\cite{Bozzi:2005wk} which resums the full DGLAP splitting functions including the off-diagonal channels in the Mellin space and performs the inverse Mellin transform numerically. This also pave the way for future applications of this resummation technique in $pA$ collisions can make phenomenological calculations in dilute dense factorizations more reliable and systematic in forward rapidity particle productions. 

\begin{acknowledgments}
We thank A. Mueller, S. Munier, F. Ringer and S.Y. Wei for useful discussion and comments, and also acknowledge the discussion with A. Stasto and D. Zaslavsky at the early stage of this work. This material is based on the work supported by the Natural Science Foundation of China (NSFC) under Grant Nos.~11575070 and by the U.S. Department of Energy, Office of Science, Office of Nuclear Physics, under contract number DE-AC02-05CH11231.
\end{acknowledgments}

\appendix
\section{An alternative derivation of the threshold resummation}
\label{a1}
We have derived the threshold resummation in an intuitive way in the context of dilute-dense factorization. In fact, following the same idea (Eq.~(3.29)) in Ref.~\cite{Becher:2006mr}, we can argue that the resummed distribution $g_{\textrm{t}} (\tau, \mu_f)$ at the scale $\mu_f$ should satisfy the simplified DGLAP evolution equation in the limit $\xi \to 1$ in order to let the cancellation of $\mu$ dependence occur in the threshold limit
\begin{equation}
\frac{d g(\tau, \mu_f)}{d\ln \mu_f} = \frac{2\alpha_s N_c}{\pi}\int_{\tau}^1 \frac{d\xi}{\xi} \mathcal{P}^{\xi\to 1}_{gg} (\xi)g\left(\tau/\xi, \mu_f\right),
\end{equation}
where $\mathcal{P}^{\xi\to 1}_{gg} (\xi) \equiv \frac{1}{(1-\xi)_+}+\beta_0\delta(1-\xi)$ which is equivalent to $\tau \to 1$ and $N\gg 1$ limits. In the Mellin space, the above equation becomes 
\begin{equation}
\frac{d g_N( \mu_f)}{d\ln \mu_f}  =\frac{2\alpha_s N_c}{\pi} \left[ -\psi (N)-\gamma_E +\beta_0\right]g_N( \mu_f).
\end{equation}
The exact solution can be written as 
\begin{equation}
g(\tau, \mu_f) =e^{(\beta_0-\gamma_E) \gamma_{\mu, \mu_f}} \int_0^1 \frac{dx}{x} g(x, \mu) \int_{\lambda -i\infty}^{\lambda +i\infty} \frac{dN}{2\pi i} \left(\frac{x}{\tau}\right)^N e^{-\psi(N) \gamma_{\mu, \mu_f}}, \label{inver}
\end{equation}
where $ \gamma_{\mu, \mu_f} \equiv N_c \int_{\mu^2}^{\mu_f^2} \frac{d \mu^{\prime 2}}{\mu^{\prime 2}} \frac{\alpha_s(\mu^{\prime 2})}{\pi}$. Although we have not been able to find an analytical form for the inverse Mellin transform in Eq.~(\ref{inver}), we can evaluate it numerically with any positive $\lambda$ when $\gamma_{\mu, \mu_f}>0$. Furthermore, if one approximates $\psi (N)$ as $\ln N$ in the large $N$ limit by using Eq.~(\ref{psiexpansion}), one can find Eq.~(\ref{inver}) becomes the results in Eq.~(\ref{thg}). It is also important to note that the same level of approximation has been made along the way when use the DGLAP equation at the end point. In addition, it is interesting to note that the exact solution in Eq.~(\ref{inver}) has the following bound 
\begin{equation}
\int_\tau^1 \frac{dx}{x}  g (x, \mu ) \Delta(\mu^2, \mu_b^2, \ln \frac{x}{\tau})  < g(\tau, \mu_f) < \int_\tau^1 \frac{dx}{x}  g (x, \mu ) \left(\frac{x}{\tau}\right)^{1/2} \Delta(\mu^2, \mu_b^2, \ln \frac{x}{\tau}), \label{bound}
\end{equation}
which has been checked numerically for sufficiently large $\tau$ when $\gamma_{\mu, \mu_f}>0$. When $\gamma_{\mu, \mu_f}\leq 0$, analytical continuation has to be applied again. In the threshold limit, the exact solution \label{inverse1} is tightly bound by Eq.~(\ref{bound}). We have also numerically checked that the above bound (especially the upper bound) usually gives excellent numerical estimate of the exact solution. Again, similar result can be obtained for quark distributions once we change to the quark splitting function accordingly. 

Let us compare our result in Eq.~(\ref{thg}) to Eq.~(3.31) in Ref.~\cite{Becher:2006mr}, which can be rewritten as (converted into gluon channel in our notation)
\begin{equation}
f_g(\tau, \mu_f) =\frac{e^{(\beta_0-\gamma_E) \gamma_{\mu, \mu_f}}}{\Gamma [ \gamma_{\mu, \mu_f}]}\int_\tau^1 dx \frac{f_g(x, \mu)}{(x -\tau)^{1-\gamma_{\mu, \mu_f}}}. \label{bnp}
\end{equation} 
First of all, our numerical evaluation of Eq.~(\ref{inver}) indicates a finite difference from Eq.~(\ref{bnp}) for $\tau<1$. Second, we find that the above equation is equivalent to Eq.~(\ref{thg}) in the limit $\tau\to 1$ with corrections of order $(1-\tau)$. Therefore, our result, which provides a slightly different analytical formulation, is complimentary to Eq.~(3.31) in Ref.~\cite{Becher:2006mr}. 

It is straightforward to generalize the above calculation up to NLO with the NLO DGLAP splitting function. In the $\xi \to 1$ limit, the $gg$ channel NLO DGLAP splitting function reads\cite{Curci:1980uw, Ellis:1991qj}
\begin{eqnarray}
\mathcal{P}^{\xi\to 1}_{gg} (\xi)& =& \frac{1}{(1-\xi)_+} \left\{ 1+\frac{\alpha_s}{2\pi}\left[N_c\left(\frac{67}{18}-\frac{\pi^2}{6}\right)-\frac{5}{9}n_f\right] \right\}  \notag \\
  && +\delta(1-\xi) \left\{\beta_0+\frac{\alpha_s}{4\pi}\left[N_c\left(\frac{8}{3}+3\zeta (3)\right)-\frac{C_F n_f}{2N_c}-\frac{2}{3}n_f\right]\right\}.
\end{eqnarray}
The NLO results for Eq.~(\ref{thg}) can be written as 
\begin{eqnarray}
 g_{\textrm{t}} (\tau, \mu_b) &=& \frac{e^{\gamma_{\beta}-\gamma_E \gamma_{\mu, b_\perp} }}{\Gamma [ \gamma_{\mu, b_\perp} ]}\int_\tau^1 \frac{dx}{x}  g (x, \mu ) \left[\ln \frac{x}{\tau} \right]_\ast^{\gamma_{\mu, b_\perp} -1}, \label{thg2}  \notag \\
\quad \textrm{with} &&  \gamma_{\mu, b_\perp} = \frac{N_c}{\pi} \int_{\mu^2}^{\mu_f^2} \frac{d \mu^{\prime 2}  \alpha_s(\mu^{\prime 2})}{\mu^{\prime 2}}\left\{ 1+\frac{\alpha_s(\mu^{\prime 2})}{2\pi}\left[N_c\left(\frac{67}{18}-\frac{\pi^2}{6}\right)-\frac{5}{9}n_f\right] \right\} ,\notag \\
&& \gamma_{\beta} = \frac{N_c}{\pi} \int_{\mu^2}^{\mu_f^2} \frac{d \mu^{\prime 2}  \alpha_s(\mu^{\prime 2})}{\mu^{\prime 2}}\left\{\beta_0+\frac{\alpha_s(\mu^{\prime 2})}{4\pi}\left[N_c\left(\frac{8}{3}+3\zeta (3)\right)-\frac{C_F n_f}{2N_c}-\frac{2}{3}n_f\right]\right\}.
\end{eqnarray}

\section{Discussion on corrections at one-loop order}
\label{a2}
Since only the dominant part of the one-loop contributions are resummed in this calculation, the difference between the exact one-loop contribution the resummed part, which vanishes in the limit $\tau \to 1$, can be computed as follows.

For the forward Higgs production in the pA collisions, the leading power one-loop contribution is proportional to 
\begin{equation}
 I^{\textrm{E}}_{\textrm{1-loop}}=\gamma_{\mu, \mu_b} \int_\tau^1 \frac{d\xi}{\xi} \mathcal{P}_{gg}(\xi) g\left(\frac{\tau}{\xi},\mu\right) = \gamma_{\mu, \mu_b} \int_\tau^1 \frac{d\xi}{\xi}\left[\frac{1}{(1-\xi)_+}-1+\frac{1-\xi}{\xi}+\xi(1-\xi)+\beta_0\delta(1-\xi)\right]g\left(\frac{\tau}{\xi},\mu\right),
\end{equation}
where the term $\frac{\alpha_s}{\pi}\frac{\pi^2}{2}N_c$ is excluded, since it is always kept in $C(\alpha_s)$. For the threshold resummation formula in Eq.~(\ref{thg}), the corresponding one-loop contribution can be obtain easily by expanding it up to first order in $\alpha_s$ (namely, $\gamma_{\mu, \mu_b}$). Alternatively, it is instructive to start from Eq.~(\ref{res2}) after expanding $\exp\left[- \gamma_{\mu, b_\perp} \ln \frac{Ne^{\gamma_E}}{e^{\beta_0}}\right] \simeq 1- \gamma_{\mu, b_\perp} \ln \frac{Ne^{\gamma_E}}{e^{\beta_0}}$. Therefore, the one-loop contribution which is resummed in Eq.~(\ref{res2}) can be explicitly computed as follows
\begin{equation}
I^{\textrm{T}1}_{\textrm{1-loop}}=-\gamma_{\mu, b_\perp}\int_\mathcal{C} \frac{dN}{2\pi i}  \tau^{-N} g_N (\mu ) \ln \frac{Ne^{\gamma_E}}{e^{\beta_0}}.
\end{equation}
Using the identity $\ln N =\lim_{\epsilon\to 0} \int_0^\infty du u^{\epsilon -1} \left(e^{-u}-e^{-uN}\right)$, one can find 
\begin{equation}
I^{\textrm{T}1}_{\textrm{1-loop}}=-\gamma_{\mu, b_\perp}\lim_{\epsilon\to 0} \int_0^1 \frac{dx}{x}g (x,\mu ) \left\{\left[\gamma_E -\beta_0 +\Gamma(\epsilon) \right]\delta(\ln x-\ln\tau)-\theta\left[\ln x-\ln\tau\right] \left(\ln\frac{x}{\tau}\right)^{\epsilon-1} \right\}. \label{stardef}
\end{equation}
Analogous to the identity for plus functions $(1-w)^{\epsilon -1} =\frac{1}{\epsilon} \delta (1-w) +\frac{1}{(1-w)_+}$, one can show the following identity, which is applied in the context of Eq.~(\ref{stardef})
\begin{eqnarray}
\left(\ln\frac{x}{\tau}\right)^{\epsilon-1} =\frac{1}{\epsilon} \delta(\ln x-\ln\tau) +\frac{1}{[\ln x-\ln\tau]_*}. 
\end{eqnarray}
The above identity can be derived in terms of $\epsilon$ expansion for any test function $f(x)$
\begin{eqnarray}
\int_\tau^1 \frac{dx}{x}f(x)\left(\ln\frac{x}{\tau}\right)^{\epsilon-1} &=& \int_\tau^1 \frac{dx}{x}\left[f(x)-f(\tau)+f(\tau)\right]\left(\ln\frac{x}{\tau}\right)^{\epsilon-1} \notag, \\
 &=&\frac{f(\tau)}{\epsilon} +\int_\tau^1 \frac{dx}{x} \frac{f(x)}{[\ln x-\ln\tau]_*}, \\
\textrm{with} \quad    \int_\tau^1 \frac{dx}{x} \frac{f(x)}{[\ln x-\ln\tau]_*}&\equiv& \int_\tau^1 \frac{dx}{x} \frac{f(x)-f(\tau)}{[\ln x-\ln\tau]}+f(\tau) \ln\ln \frac{1}{\tau}.
\end{eqnarray}
Therefore, it is straightforward to find that the final result is finite and it reads
\begin{equation}
I^{\textrm{T}1}_{\textrm{1-loop}}=\gamma_{\mu, b_\perp} \int_\tau^1 \frac{dx}{x}g (x,\mu ) \left[\frac{1}{[\ln x -\ln \tau]_*}+\beta_0\delta(\ln x-\ln\tau)\right].
\end{equation}
Then, the contributions which are not resummed in this approach can be cast into
\begin{eqnarray}
 I^{\textrm{E}}_{\textrm{1-loop}} -I^{\textrm{T}1}_{\textrm{1-loop}} = \gamma_{\mu, b_\perp} \int_\tau^1 \frac{dx}{x} g(x,\mu)  \left[\frac{1}{(1-\xi)_+}-\frac{1}{[\ln1/\xi]_*} -1 +\frac{1-\xi}{\xi}+\xi(1-\xi)\right]_{\xi=\frac{\tau}{x}}. 
\end{eqnarray}
Due to the cancellation between the plus-function and the star-distribution at the end point where $\xi=1$ or $x=\tau$, we can find that the difference is no longer singular and the corresponding remaining contribution vanishes in the limit $\tau\to 1$ for fixed $\gamma_{\mu, b_\perp} $. As a comparison, let us consider the resummation formula in Eq.~(\ref{bnp}) derived in Ref.~\cite{Becher:2006mr}. Similarly, one can find that the corresponding one-loop contribution and finite differences read
\begin{eqnarray}
&& I^{\textrm{T}2}_{\textrm{1-loop}}=\gamma_{\mu, b_\perp}   \int_\tau^1 dx  \frac{g (x,\mu )-g(\tau,\mu)}{x-\tau}+ \gamma_{\mu, b_\perp} g(\tau,\mu) \ln (1-\tau)+\gamma_{\mu, b_\perp} \beta_0 g(\tau,\mu), \\
&&  I^{\textrm{E}}_{\textrm{1-loop}} -I^{\textrm{T}2}_{\textrm{1-loop}}=\gamma_{\mu, b_\perp} \int_\tau^1 \frac{dx}{x} g(x,\mu) \left[\delta (1-\xi) \ln \frac{1}{\tau}-1 +\frac{1-\xi}{\xi}+\xi(1-\xi)\right]_{\xi=\frac{\tau}{x}}.
\end{eqnarray}
The above difference also vanishes when $\tau \to 1$. In addition, one also needs to note that there is off-diagonal contribution from quark to gluon splittings, which is again of order $\alpha_s (1-\tau)$.

\end{document}